\documentclass[doublecol]{epl2} 
\usepackage{color}
\usepackage{epstopdf}

\definecolor{red}{rgb}{1,0,0}
\definecolor{orange}{rgb}{1,0.5,0}
\definecolor{blue}{rgb}{0,0,1}
\definecolor{green}{rgb}{0,1,0}

\title{Wave turbulence in vibrating plates : the effect of damping}
\shorttitle{Wave turbulence in vibrating plates : the effect of damping} 

\author{T. Humbert\inst{1,2} \and O. Cadot\inst{2} \and G. D\"uring\inst{3} \and C. Josserand\inst{1} \and S. Rica$^\dag$\inst{4} \and C. Touz\'e\inst{2} }
\shortauthor{T. Humbert \etal}

\institute{                    
  \inst{1} Institut D'Alembert, UMR 7190 CNRS-UPMC, 4 place Jussieu, 75005 Paris, France.\\
  \inst{2} Unit\'e de M\'ecanique (UME), ENSTA ParisTech, 828 Bd des Mar\'echaux, 91762 Palaiseau Cedex, France \\
 \inst{3} Physics Department, New-York University, 4 Washington Place, New York, NY 10003, USA.\\
\inst{4} Facultad de Ingenier\'ia y Ciencias, Universidad Adolfo Ib\'anez, Avda. Diagonal las Torres 2640, Penalolen, Santiago, Chile\\
}

\abstract{
The effect of damping in the wave turbulence regime for thin vibrating plates is studied. An experimental method, allowing measurements of dissipation in the system at all scales, is first introduced. Practical experimental devices for increasing the dissipation are used. The main observable consequence of increasing the damping is a significant modification in the slope of the power spectral density, so that the observed power laws are not in a pure inertial regime. However, the system still displays a turbulent behavior with a cut-off frequency that is determined by the injected power which does not depend on damping. By using the measured damping power-law in numerical simulations, similar conclusions are drawn out.}

\begin{document}

\maketitle

\section{Introduction}

Wave (or weak) turbulence theory (WTT) aims at describing the long time behavior of weakly non linear systems with energy exchanges between scales. It predicts for long time broadband Kolmogorov-Zakharov spectra, by analogy with hydrodynamic turbulence~\cite{Zakharov_91,NewRum11,Naza}. A large number of situations have been studied over the years starting from the initial context of water waves~\cite{Hass62,ZakFil67,Janssen_03, Falcon_98}, to nonlinear optics~\cite{Dyachenko_91} or Alfven Waves in plasmas~\cite{Galtier_00} for instance.\\
\indent Wave turbulence for elastic vibrating plates has been investigated theoretically in 2006~\cite{Josserand_06}, rapidly followed by two experimental works~\cite{Boudaoud_08,Mordant_08}. The theoretical analysis considers the dynamics in the framework of the von K{\'a}rm{\'a}n equations. For a thin plate of thickness $h$, Poisson ratio $\nu$, density $\rho$ and Young's modulus $E$, it yields~\cite{landau}
\begin{eqnarray}
\label{eq.1}
\rho h\frac{\partial^2\zeta}{\partial t^2}&=&-\frac{Eh^3}{12(1-\nu^2)}\Delta^2\zeta+{\mathcal L}(\chi,\zeta),\\
\Delta^2\chi&=&-\frac{Eh}{2}{\mathcal L}(\zeta,\zeta),
\end{eqnarray}
where $\zeta$ is the transverse displacement and $\chi$ the Airy stress function. The operator ${\mathcal L}$ is bilinear symmetric, and reads in cartesian coordinates: ${\mathcal L}(f,g)=f_{xx}g_{yy}+f_{yy}g_{xx}-2f_{xy}g_{xy}$. Such dynamics exhibits dispersive waves, following the dispersion relation
\begin{equation}
\label{eq.2}
\omega_{\mathbf k}= h c | {\mathbf k} |^2
\end{equation}
involving the bulk sound velocity
\begin{equation}
c=\sqrt{\frac{E}{12(1-\nu^2)\rho}}.
\end{equation}
\noindent Then, developing the usual wave turbulence analysis, non equilibrium solutions of the kinetic equation have been found in~\cite{Josserand_06}, the so-called Kolmogorov-Zakharov spectra characterized by the relation between the power spectral density $P_v$ of the velocity $v$ and the frequency $f= \omega/(2 \pi)$:
\begin{equation}
\label{eq.3}
P_v \propto  P^\frac{1}{3} {\rm log}^\frac{1}{3} \left(\frac{f_c}{f} \right)
\end{equation}
where $P$ is the energy flux which is transferred along the cascade until it is dissipated near $f_c$, the cut-off frequency of the spectrum. Such spectra were also observed in 
numerical simulations~\cite{Josserand_06} by injecting energy at small frequencies (large scales) and dissipating it at high 
frequencies (small scales with $f_c$ constant).\\
Surprisingly, two independent experiments, performed soon after on thin elastic plates~\cite{Mordant_08, Boudaoud_08}, did not replicate these theoretical (and numerical) predicted spectra. Both experiments have measured similar turbulent spectra, following
\begin{equation}
\label{eq.4}
P_v \propto \epsilon_I^\frac{1}{2} \left(\frac{f}{f_c} \right)^{-\frac{1}{2}} \quad {\rm with} \quad f_c\propto \epsilon_I^{1/3}.
\end{equation}
where $\epsilon_I$ is the mean injected power, a quantity related to the energy flux, but slightly different. Indeed, some of the injected 
energy can be dissipated at large scale by the plate modes without entering the cascade process.
Beside a different power-like behavior of the spectra ($f^0$ compared to $f^{-0.5}$), the dependence on $\epsilon_I$ is puzzling since it may indicate that the expected four-waves resonance is failing. In fact, since the critical frequency $f_c$ varies with the injecting power, it leads to an overall dependence of the spectra with the injected power  $P_v \propto  \epsilon_I^\frac{2}{3}$.

Usually, four main mechanisms are invoked to explain the discrepancies between theoretical and experimental spectra: 1) the finite size 
of the experimental system~\cite{Karta_94,Miquel_11}; 2) an incorrect separation of the linear and nonlinear time 
scales~\cite{Miquel_12}; 3) the influence of the strongly non linear regime~\cite{Yokov}; 4) the dissipation that can invade the transparency window where the cascade dynamics holds. 
Concerning the latter phenomenum, weak turbulence theory requires dissipative scales to be widely separated from forcing scales to allow the conservation of the energy flux through the cascade. In solid plates, the dissipation has different origins and is in fact present at every scales~\cite{Chaigne_01}, so that this property is highly questionable. The goal of this paper is therefore to quantify both experimentally and numerically the influence of the real dissipation on the turbulent spectra.\\

\section{Experiments}

\begin{figure}
\onefigure[width=0.469\textwidth]{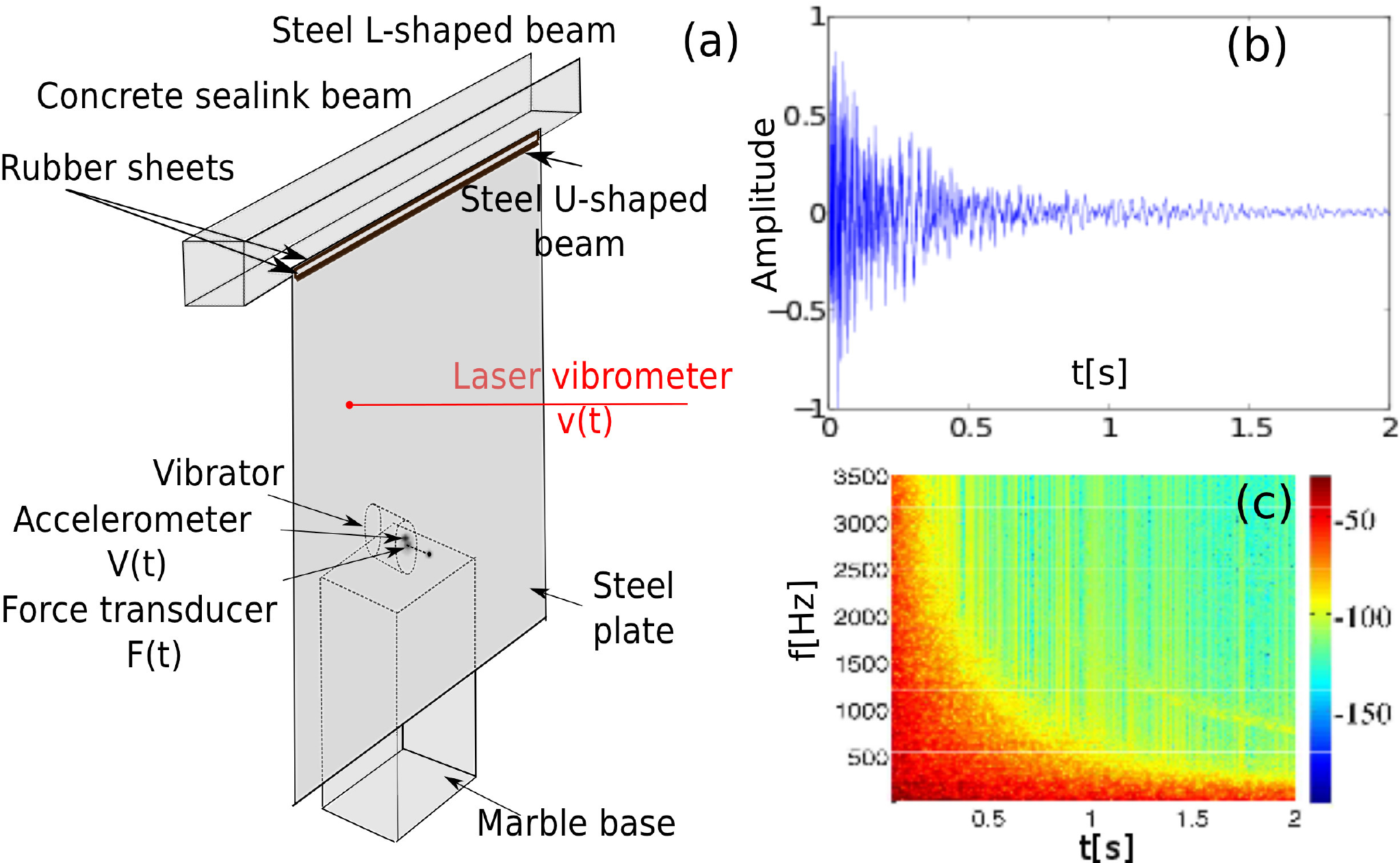}
\caption{ (a) Experimental set-up: the steel plate is clamped at its top and the vibrations are enforced by a shaker. A force transducer and an 
accelerometer measure force and velocity at the injection; a laser vibrometer records the transverse velocity at a given position on the plate.
(b) Amplitude of the impulse response as a function of time for the plate in configuration $2SP$ (see text). (c) Spectrogram (in dB) of the impulse 
response (b) as a function of time $t$ and frequency $f$.}
\label{schema}
\end{figure}

The plate is made of steel with $\rho=7800 \, {\rm kg \cdot m}^{-3}$, $E=210$ GPa and $\nu=0.3$, giving $c=1570 \, {\rm m \cdot s}^{-1}$. The lateral dimensions are $2 \times 1\,{\rm m}^2$, and the thickness of the plate is $h=0.5$mm, so that the lowest fundamental vibrating frequency of the plate is of the order of $1$ Hz. It is hanged under its own weight and clamped on the top side as described in fig.~\ref{schema}. A LDS shaker V455MS is placed at mid-width of the plate and $62$cm away from the bottom side. A force transducer Bruel \& Kjaer Type 8230-002 is mounted between the shaker and the plate to measure the force applied from the shaker to the plate $F(t)$. An accelerometer Bruel \& Kjaer Type 4517 mounted on the shaker gives the injection speed $V(t)$. Both $V(t)$ and $F(t)$ are used to deduce the mean injected power

\begin{equation}
\epsilon_I = \frac{\left< F(t) \cdot V(t) \right>}{\rho S},
\end{equation}
where the brackets denote a temporal mean and $S$ the surface of the plate. The plate is set into a turbulent regime with a 
sinusoidal forcing at frequency $f_0=30$Hz which corresponds to a low frequency mode of the plate.\\
A Polytec laser vibrometer OFV 056 measures the transverse velocity $v(t)$ at a point located $1$m from the bottom of the plate 
and $40$cm from the left edge (see fig.~\ref{schema}). The data are sampled at $22$kHz. The velocity power spectra are 
time-averaged on windows of $1$s over $180$s, so that the spectral resolution is $1$Hz.\\

\indent The natural damping of the plate (configuration $N$) is increased using two different techniques. The first one 
generates a homogeneous damping by painting one (configuration $1SP$) or the two ($2SP$) sides of the plate. The 
second method consists in adding dampers on all free edges of the natural plate in order to attenuate the reflected waves (configuration 
ED). These dampers have cylindrical shape with diameters of $1.7$cm and are commercially used for thermic isolation of hydraulic 
pipes.\\
\begin{figure}
\onefigure[width=0.43\textwidth]{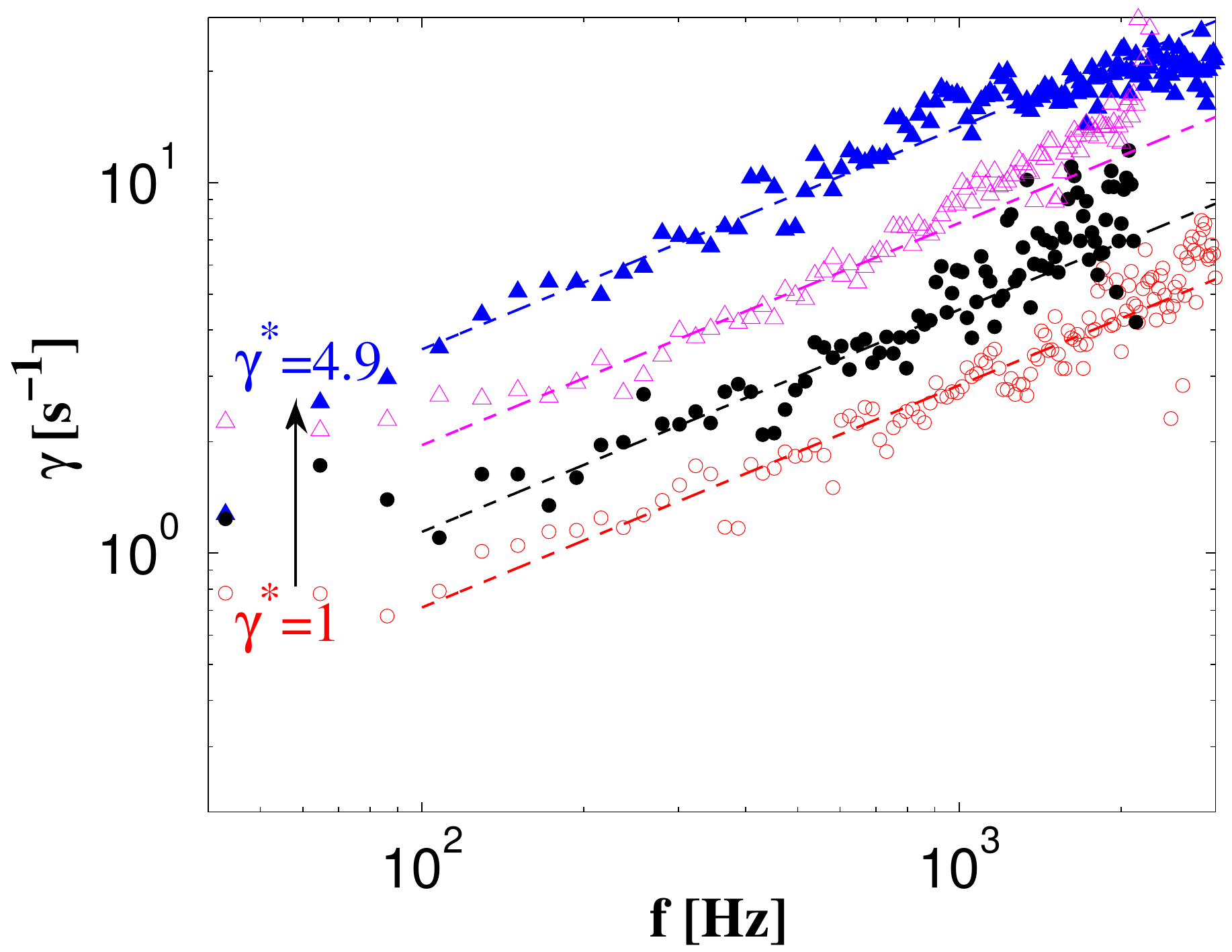}
\caption{Evolution of the damping factor $\gamma$ as a function of the frequency. Empty red circles: configuration $N$, filled black 
circles: $1SP$, empty magenta triangles : $2SP$, filled blue triangles: $ED$. Dashed lines + equations: fitted power laws for 
numerical simulations.}
\label{Amortissement}
\end{figure}
\indent Each configuration is characterized by the measurements of the attenuation coefficient in the linear regime. Due to the very high modal density of the experimental plate, a measurement based on the impulse response is the most appropriate. However, due to the large size of the system, 
the classical technique using an impact hammer leads to a poor signal-to-noise ratio. \\
\indent The impulse response $y_I(t)$ is
constructed here from the velocity response $y(t)$ to a broadband excitation signal $x(t)$, recorded in one point and using the inverse filter $x(T-t)E(t)$,
\begin{equation}
y_I=y(t)\otimes x(T-t)E(t).
\end{equation}
$E(t)$ has to fulfill the relation $x(t)\otimes x(T-t)E(t)=\delta$, where T is the signal length, $\delta$ the Dirac delta function and $\otimes$ the convolution product.
The method, developed in~\cite{Farina_00} for room acoustics and adapted to reverberated plates in~\cite{Arcas_09}, 
uses a logarithmic sine sweep
\begin{equation}
x(t)=\sin[\frac{2\pi f_1T}{\ln(f_2/f_1)}(e^{\frac{t}{T}\ln(f_2/f_1)}-1)],
\end{equation}
\noindent where $f_1$ and $f_2$ are respectively the smallest and largest frequencies. In that case
\begin{equation}
E(t)=e^{\frac{t}{T}\ln(f_2/f_1)(\frac{-6}{\log_{10}(2)})}.
\end{equation}
The logarithmic sine sweep is particularly interesting for nonlinear systems because it allows for a clear distinction between the linear and the
nonlinear components. In the resulting signal $y_I(t)$, the linear response starts at $t=T$ whereas the contributions of 
each harmonic coming from the nonlinear distortion appears before $t=T$~\cite{Farina_00,Arcas_09}. In our case, even for the lowest amplitude of $x(t)$, distortions are always observable.
We have checked that the linear impulse response does not depend on the amplitude of the exciting signal, confirming the accuracy 
of the technique. Practically, we take $f_1=20 \,{\rm Hz}$ and $f_2=3 \, {\rm kHz}$ and the signal is produced at the analog output of the acquisition board with
a sample rate of $22 \, {\rm kHz}$. The velocity measured by the vibrometer is simultaneously recorded at the same sample rate. 
Fig.~\ref{schema}(b)  displays the impulse response of the configuration $2SP$. As it can be seen in the spectrogram (fig.~\ref{schema}(c)), the energy decreases with time, faster for the high than for the low frequencies. This energy decrease can fitted for each frequency by an exponential law $\exp^{-\gamma t}$ defining the damping factor $\gamma$. Finally, we have checked that the technique proposed in~\cite{Miquel_11} reproduces satisfactorily the damping factor measured with our method which exhibits however a greater accuracy since it discriminates directly the linear damping from non-linear effects.

\begin{figure}[t]
\onefigure[width=0.38\textwidth]{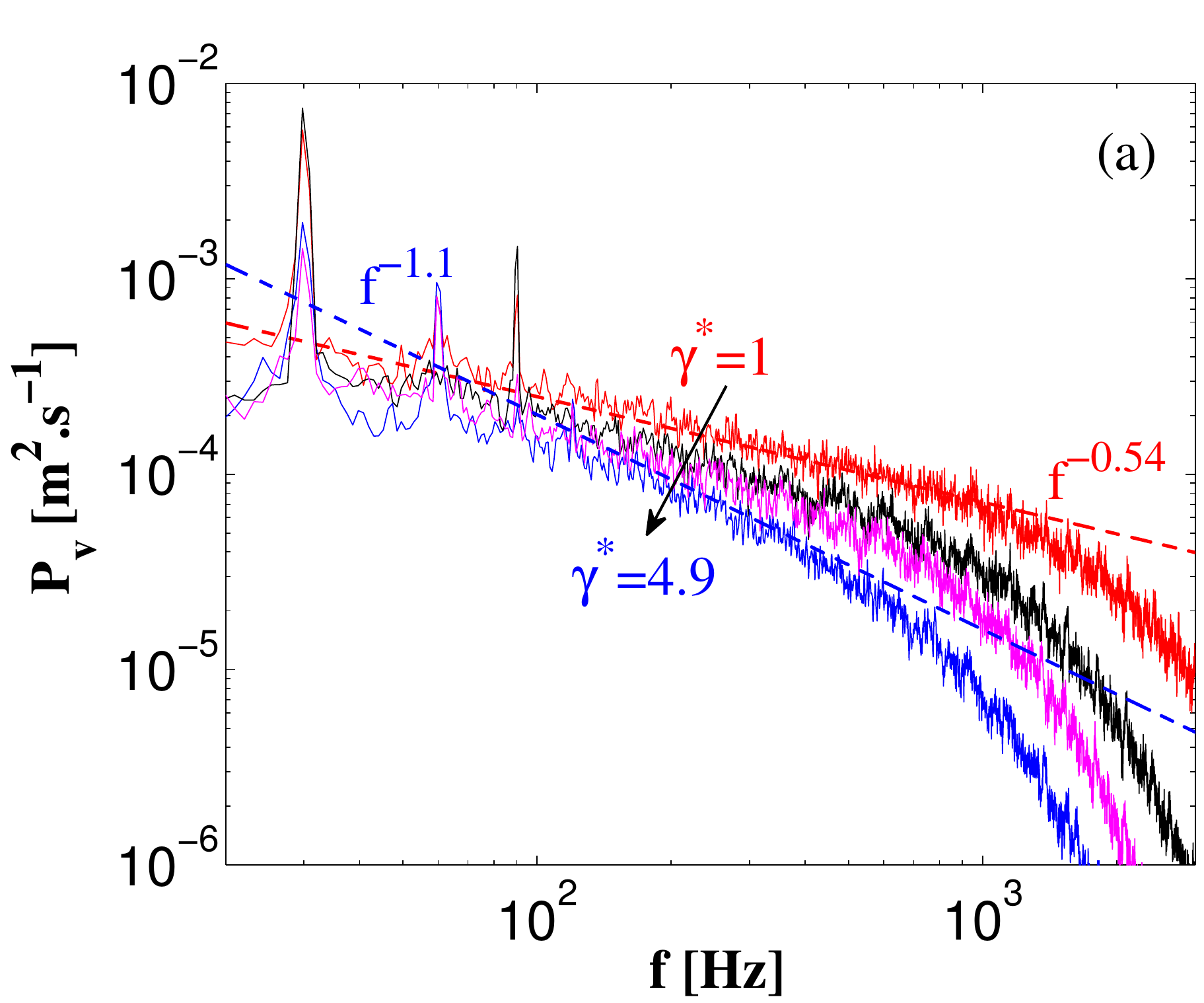} 
\onefigure[width=0.38\textwidth]{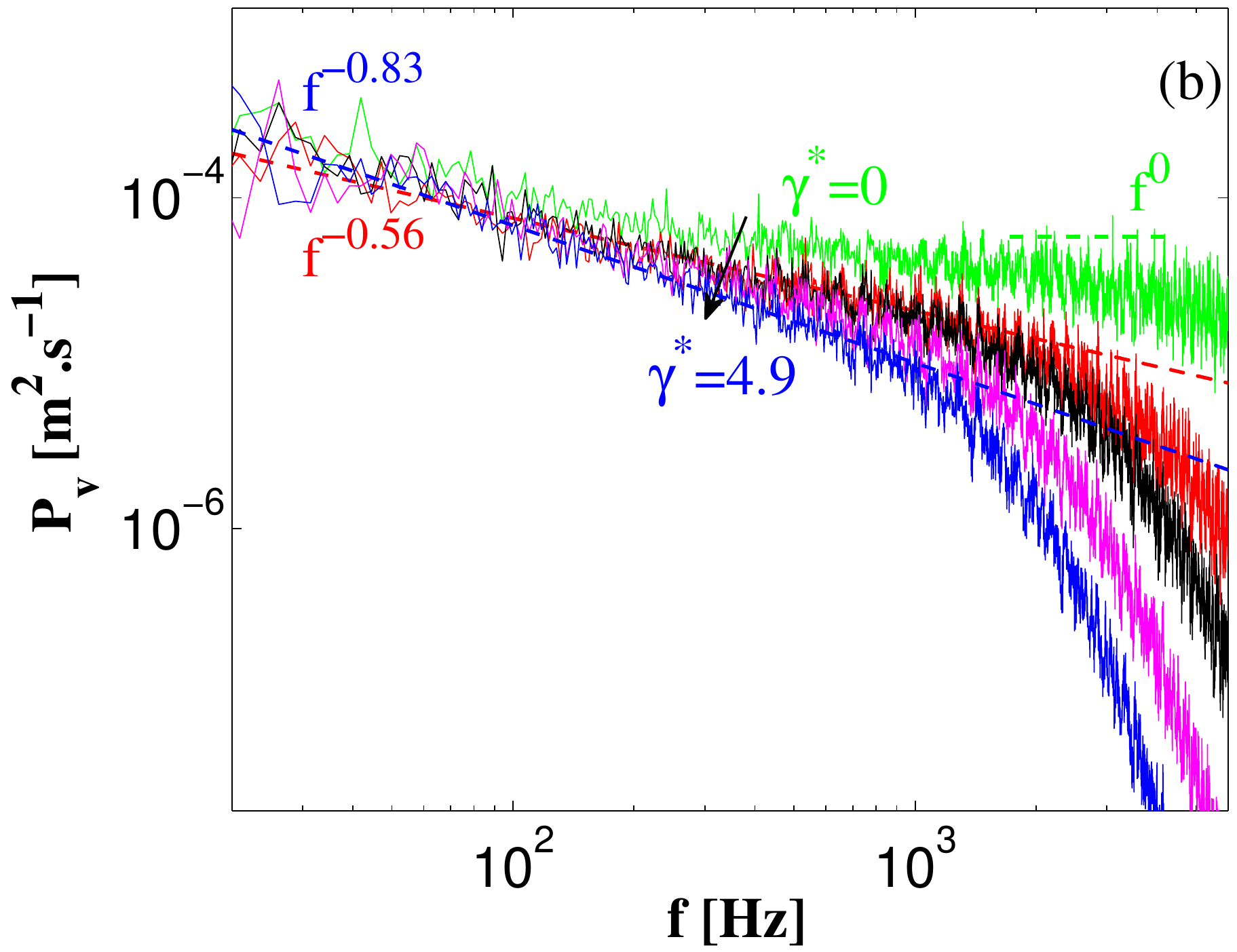} 
\caption{Power spectral density of the transverse velocity for the four configurations. Red dashed line + equation: smallest slope. Blue dashed line + equation: largest slope. (a) Experiments: for $\gamma_*=1$ (red) $\epsilon_I=0.56 \times 10^{-3} \, {\rm m^3 \cdot s^{-3}}$, for $\gamma_*=1.6$ (black) $\epsilon_I=0.54 \times10^{-3} \, {\rm m^3 \cdot s^{-3}}$, for $\gamma_*=3.1$ (magenta) $\epsilon_I=0.52 \times10^{-3} \, {\rm m^3 \cdot s^{-3}}$, and for $\gamma_*=4.9$ (blue) $\epsilon_I=0.48 \times 10^{-3} \, {\rm m^3 \cdot s^{-3}}$. (b) Numerics :  for $\gamma_*=0$ (green) $\epsilon_I=0.057\, {\rm m^3 \cdot s^{-3}}$, other cases $\epsilon_I=0.024 \, {\rm m^3\cdot s^{-3}}$ }
\label{Spectres}
\end{figure}

Fig.~\ref{Amortissement} displays the evolution of the damping factors $\gamma(f)$ for the four experimental configurations. Interestingly, despite the 
different attenuation sources, the damping factors exhibit always the same qualitative behavior which can be characterized by a power law dependence of 
the attenuation on the frequency with an exponent close to $0.6\pm 0.05$. The damping is due to very different physical origins: thermoelasticity, 
acoustic radiation and dissipation at the boundaries in particular. In our case the coincidence frequency for acoustic radiation can be estimated in the vicinity of 20 kHz~\cite{Chaigne_01, Arcas_09} and is thus not relevant for our frequency range. Thermal effects can be described by Zener model~\cite{Zener_38} which leads to a frequency-independent~\cite{landau, Chaigne_01, Arcas_09} small value for the damping factor, roughly estimated at $0.8 s^{-1}$ for our plate. A part
of the energy is also dissipated at the clamped
boundary where two pieces of rubber are inserted in order
to avoid a metallic contact between the plate and the
clamp. A physical model for the dissipation should also account 
for the viscoelastic behavior of the rubber that have
to be measured independently and is beyond the scope of
the present study. In the remainder and for further analysis, we will use the robust relation found in the experiments:
\begin{equation}
\gamma(f)=\alpha f^{0.6},
\label{fit_amort}
\end{equation}
where $\alpha$ will be fitted for each experimental configuration. The four different plates will be characterized by their relative coefficient $\gamma_{\ast}$, ratio of the damping coefficient with that of the natural case $\alpha_N$
\begin{equation}
\gamma_*=\frac{\alpha}{\alpha_N}.
\end{equation}
The damping coefficients $\gamma_*$ vary between $1$ and $5$ (see Table~\ref{tab.1}). Note that the fitted law (\ref{fit_amort}) is particularly good for high frequencies while it shows important 
discrepancies with the real dissipation at small frequencies corresponding to the plate eigen-modes (frequencies corresponding to 
wave-numbers of the order of the plate dimensions).
\begin{table}[h!]
\caption{$\gamma_*$}
\label{tab.1}
\begin{center}
\begin{tabular}{lcccc}
 Config & $N$ & $1SP$ &$2SP$ & $ED$ \\
$\gamma_*$ & 1 & 1.6 & 3.1 & 4.9
\end{tabular}
\end{center}
\end{table} 

From these damping measurements it is possible to estimate the power dissipated by the plate fluctuations at the location of the velocity measurements. Actually, this dissipated power is simply obtained from the experimental power spectra $P_v(f)$:
\begin{equation}
\label{eq_epsd}
\epsilon_D=h\int_0^\infty \gamma(f) P_v(f)df,
\end{equation}
using the fitted law (\ref{fit_amort}) for $\gamma(f)$. In the experiments, $\epsilon_D$ and $\epsilon_I$ are found to be proportional with $\epsilon_D=0.44\epsilon_I$, showing that a fraction of the injected power does not go into the cascade. This misfit can be ascribed to the non-homogeneity of the experimental turbulence, but also to the fitted law (\ref{fit_amort}) that underestimates the amount of energy dissipated by the very first modes of the plate. Even if $\epsilon_D$ represents better what is dissipated by the cascade, $\epsilon_I$ will be used as a control parameter in the following for the sake of coherence with previous studies~\cite{Mordant_08, Boudaoud_08}.

\section{Numerics}
Numerical simulations of the von K{\'a}rm{\'a}n plate equations are performed using the same pseudo-spectral method than in~\cite{Josserand_06}. 
Within this framework, it is straightforward to inject energy at controlled scales and to mimic the measured experimental 
dissipation. Formally, we write the following set of dynamical equations in the Fourier space:
\begin{equation}
\label{numfou}
\rho h\frac{\partial^2\zeta_{\bm k}}{\partial t^2}=-\frac{Eh^3 k^4}{12(1-\nu^2)}\zeta_{\bm k}+{\mathcal L}(\chi,\zeta)_{\bm _k}+f^i_{\bm k}-\rho h \gamma^d_{\bm k} \frac{\partial \zeta_{\bm k}}{\partial t},\\
\end{equation}
\begin{equation}
k^4\chi_{\bm k}=-\frac{Eh}{2}{\mathcal L}(\zeta,\zeta)_{\bm k},
\end{equation}
where the Fourier transform is indexed by its wave-vector ${\bm k}$ subscript.  $f^i_{\bm k}$ stands for the injection term in the Fourier space, while the dissipation is provided by the damping term $\rho h \gamma^d_{\bm k} \frac{\partial \zeta}{\partial t}$. Finally ${\mathcal L}(\chi,\zeta)_{\bm k}$ and ${\mathcal L}(\zeta,\zeta)_{\bm k}$ denote the Fourier transform of the nonlinear terms. To describe qualitatively the experimental injection at low frequency, the term $f^i_{\bm k}$ is taken as a random field in a range of 
wave number $k_{min} \le |k| \le k_{max}$, which corresponds to random excitation of the plate in the pulsation range
$ \omega_{min}=hck_{min}^2 \le \omega \le  \omega_{max}=hck_{max}^2$ using the relationship (\ref{eq.2}). $\gamma^d_{\bm k}$ is directly deduced 
from the experimental fitted law (\ref{fit_amort}), giving:
\begin{equation}
\gamma^d_{\bm k}= \alpha \left(\frac{hck^2}{2 \pi}\right)^{0.6}=\alpha \left(\frac{hc}{2 \pi}\right)^{0.6} k^{1.2}.
\end{equation}

We simulate in the numerics a $1$m$\times 1$m plate with the same mechanical properties than the experimental plate, using periodic boundary conditions.  $128 \times 128$ spatial modes are solved, in good quantitative agreement with the range
of frequency spanned in the experiments. The same attenuation coefficients than for the experiments are taken and the injection is
made in the frequency range $[5:35]$Hz. Finally, we consider also the ideal case for wave turbulence $\gamma_0=0$ where the dissipation is present at small scale only as in~\cite{Josserand_06} (formally only above a critical frequency $f_c$ which corresponds to the cut-off length scale $\lambda_c \sim 1$cm). 

\section{Results}
Experimental and numerical results are presented simultaneously for the four different configurations. The case $\gamma_*=0$ is shown numerically as a guideline of the WTT predictions.

\indent Fig.~\ref{Spectres}(a) displays the experimental power spectral densities of the normal velocity for similar injected powers. All these spectra 
exhibit turbulent-like behavior since a large range of frequencies is filled,  showing a cascade process from the large to the small scales. They all behave roughly as power laws in the cascade regime with 
frequency exponents that become clearly smaller as $\gamma_*$ increases. For the natural plate the exponent ($-0.5$) is consistent with the previous 
results~\cite{Boudaoud_08,Mordant_08}, while for the most damped plate, the exponent is almost twice this value. Fig.~\ref{Spectres}(b) shows the 
power 
spectral density of the normal velocity obtained by numerical simulations for similar injected power. The same behavior is observed. The ideal case $\gamma_*=0$, showing the almost flat spectrum (\ref{eq.3}), is drawn for comparison.\\

\begin{figure}
\onefigure[width=0.43\textwidth]{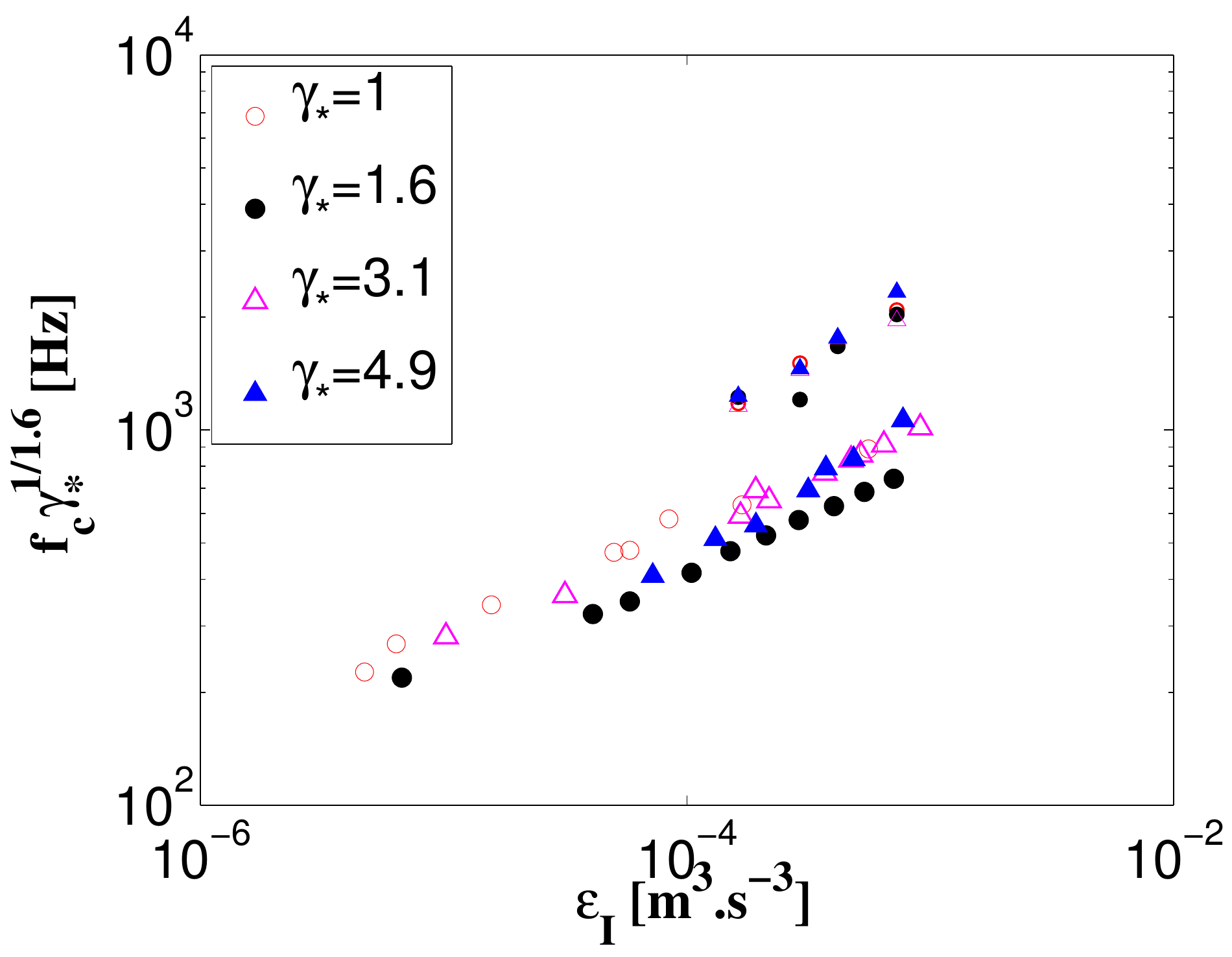}
\caption{Scaling law test for the energy budget of the cascade given in eq.~\ref{scaling_law} (see text). Large markers: experiments. Small markers: numerics.}
\label{ExpNum}
\end{figure}

\section{Discussion}

\indent These first results clearly highlight that the slope of the turbulent power spectra in vibrating plates depends strongly on the damping, indicating that it must be retained as a pertinent feature to explain the difference between theory and experiments. Moreover, 
one can argue that no inertial range (or transparency window) exists in these turbulent regimes. This can be seen in the dissipation spectrum, defined as
\begin{equation}
\gamma(f)P_v(f) =\alpha f^{0.6}P_v(f)
\end{equation}
which spreads over all the frequencies in the cascade region, since $P_v(f)\propto f^{-\beta}$ with $\beta$ varying between $0.5$ and $1.1$. For the highest attenuation studied here ($\beta=1.1$), the dissipation spectrum is even higher at large scale than at short scale, in total contradiction with the wave turbulence framework. 
It is therefore difficult to draw conclusions from the experimental observation using the general properties of wave turbulence, obtained in a conservative framework.\\
In particular, it has been shown in previous studies~\cite{Mordant_08, Boudaoud_08}, that the experimental spectra can be rescaled on a single curve as the injected power varies, by using the relation: $P_v(\frac{f}{f_c})/\sqrt{\epsilon_I}$, where $f_c$ is the cut-off frequency defined by:
\begin{equation}
f_c=\frac{\int^\infty_{f_0}P_v(f)fdf}{\int^\infty_{f_0}P_v(f)df}.
\end{equation}
This result is in apparent contradiction with the WTT prediction (\ref{eq.3}) based on four wave resonances since the power law dependence (regardless 
on the variation of $f_c$ with $\epsilon_I$) in the injected power suggests that three waves resonances are dominant. 
Alternatively, we propose here to extract the self-similar properties of the spectra directly from the cut-off frequency.

Indeed, the physical interpretation of the cut-off frequency can be clarified~\cite{Mordant_08, Boudaoud_08}, assuming that the cascade stops because the injected power has been completely dissipated by all the excited modes, yielding
\begin{equation}
\epsilon_D=h\int_0^{\infty}\gamma(f) P_v(f)df\simeq h\int_0^{f_c}\gamma(f) P_v(f)df
\end{equation}
Using the expected self-similar scaling $P_v(f)\propto\epsilon_I^{\lambda}(f/f_c)^{-\beta}$ with $\lambda$ unknown, the energy budget of the cascade becomes
\begin{equation}
\epsilon_D\propto\gamma_* f_c^{1.6}\epsilon_I^{\lambda}.
\end{equation}
Note that this relation does not depend on the slope of the spectra $\beta$. Finally, since $\epsilon_I$ is proportional to $\epsilon_D$, we obtain the following expression for the cut-off frequency
\begin{equation}
\label{scaling_law}
\gamma_*^{1/1.6}f_c\propto \epsilon_I^\frac{1-\lambda}{1.6}.
\end{equation}
Fig.~\ref{ExpNum} shows the cut-off frequency as a function of the injected power for the different damping configurations both in the experiments and
the numerics. The vertical shift between the numerics and the experiments can be simply explained by the ratio in the experiments
between $\epsilon_I$ and $\epsilon_D$, which generate a systematic translation of the experimental data (logarithmic scale). $\lambda$ can thus be computed for each different configuration: it varies from $0.36$ to $0.57$ in experiments and from $0.33$ to $0.39$ in the numerics, depending on the damping coefficient. Note that $\lambda$ evolves between $1/3$ and $1/2$, the two relevant values in the WTT for four-waves and three-waves resonances respectively.

\begin{figure}[t]
\onefigure[width=0.43\textwidth]{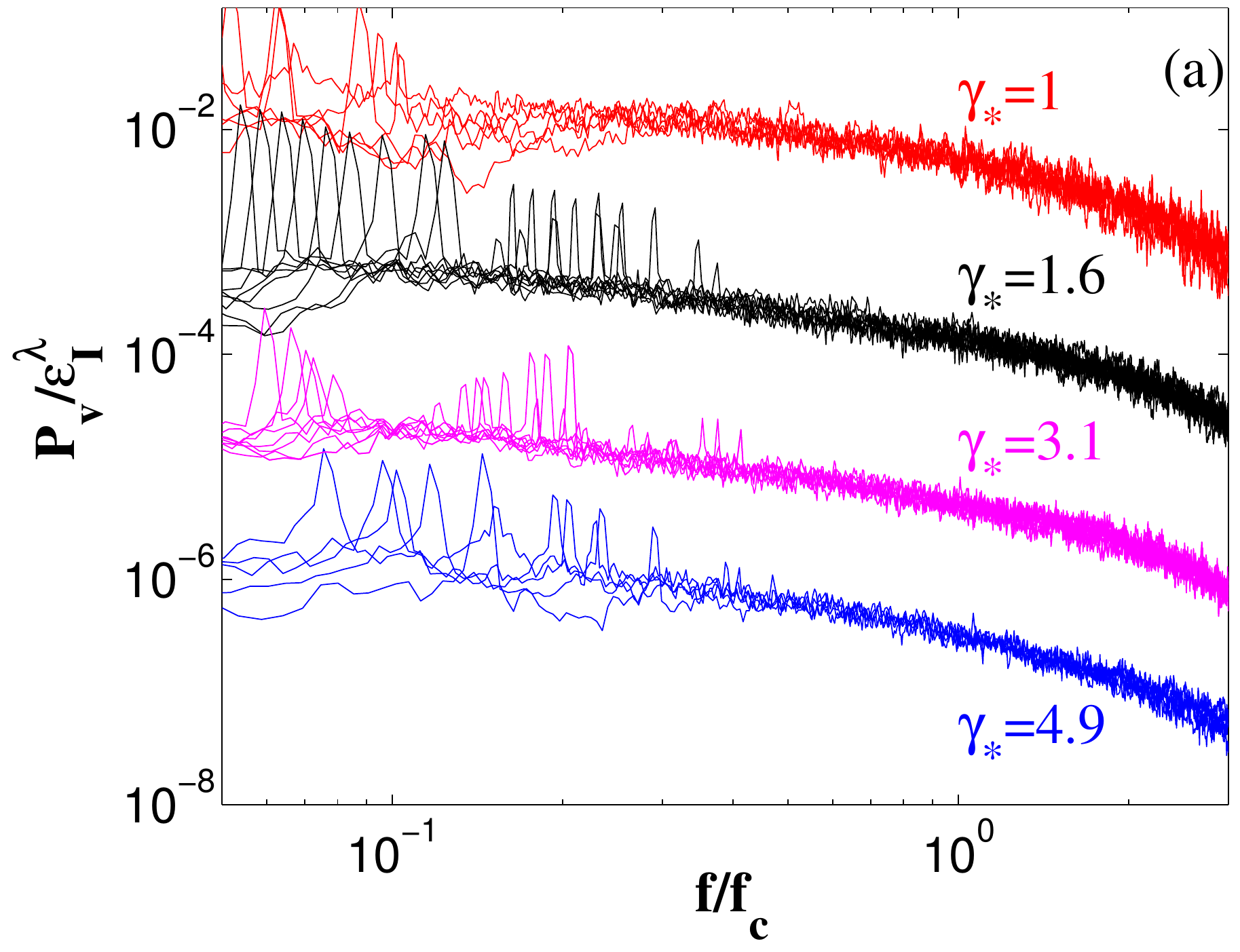} 
\onefigure[width=0.43\textwidth]{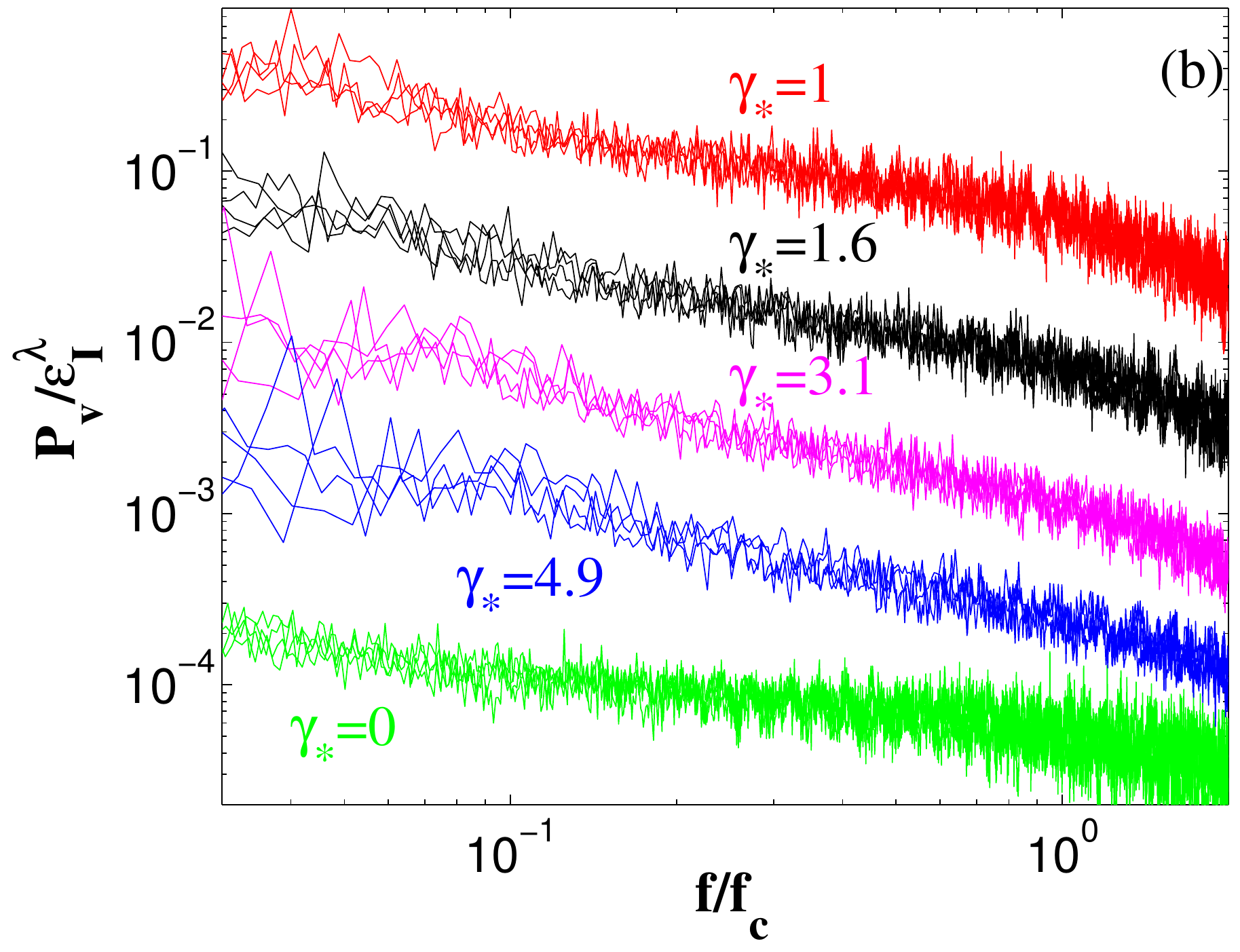} 
\caption{Power spectral density of the transverse velocity as a function of the rescaled frequency $f/f_c$ for several injected powers. Amplitude scaled by the power $\lambda$ of the injected power. Green: $\gamma_*=0$, red: $\gamma_*=1$, black: $\gamma_*=1.6$, magenta: $\gamma_*=3.1$, blue: $\gamma_*=4.9$. The curves for the different $\gamma_*$ are switched for readability. (a) Experiments. (b) Numerics.}
\label{Spectresadim}
\end{figure}

Using these measured values of $\lambda$, one can test the self-similar scaling for $P_v(f)$ as shown in fig.~\ref{Spectresadim}. Both in the 
experimental and numerical cases and for the different damping coefficients, the collapses of the curves are very good, suggesting that the injected 
power dependance is also depending on the damping, rather than on the underlying wave-resonance process.

\begin{figure}
\onefigure[width=0.43\textwidth]{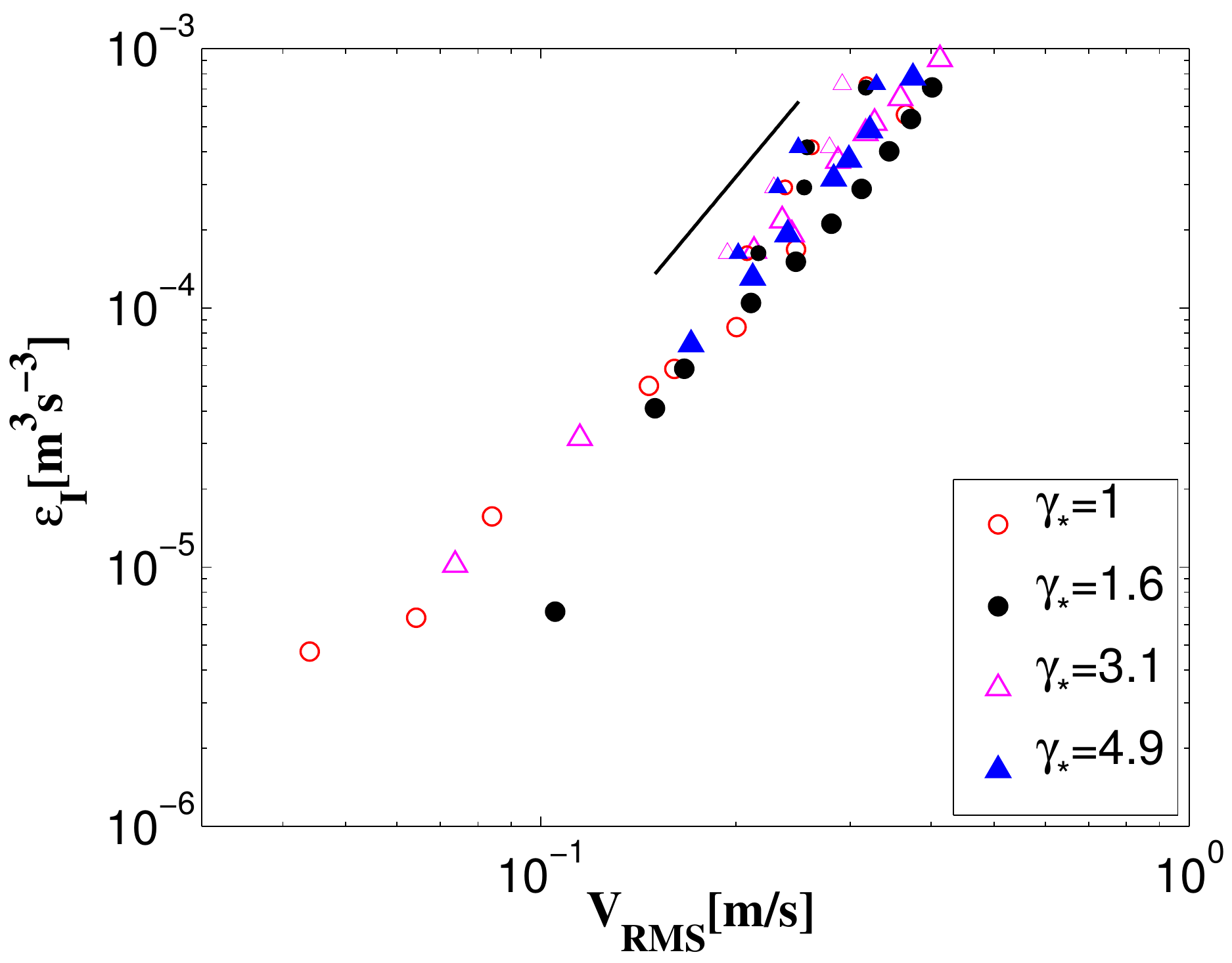}
\caption{Evolution of the injected power $\epsilon_I$ as a function of the root mean square of the injection velocity $V_{RMS}$. Large markers: experiments, small markers: numerics. Continuous line : $\epsilon_I\propto V_{RMS}^3$.}
\label{Exp5}
\end{figure}

Finally, this analysis definitely shows that a cascade process of constant energy flux is not at hand in turbulent plate vibrations. On the contrary, energy 
flux decreases all along the energy transfer towards small scales, defining both the slope and the cut-off frequency as a function of the amount of 
damping. Note however that the system is truly {\it turbulent}, with cascades displaying power laws that end when all the excited modes have dissipated 
the feeding energy. The turbulent dynamics is also testified by the relation between the injected power and the injected velocity (fig.~\ref{Exp5}), 
characterized experimentally by the RMS value of the velocity at the injection point and in the numerics by the square root of the integral over the injection scales of $P_v$. The figure shows that the injected velocity depends only on the injected power for the different damping configurations yielding 
\begin{equation}
\epsilon_I\propto V_{RMS}^3.
\end{equation} 
Thus, the mechanism of power injection is inertial as it is observed in hydrodynamic turbulence when varying the viscosity~\cite{Cadot_97}. 

\section{Conclusion}

The effect of damping on the turbulent behavior of vibrating plates has been investigated both experimentally and numerically. The energy spectra 
exhibit power law like behaviors with exponents that decrease with increasing damping.
Even though, the presence of a turbulent regime in which a cascade process is involved, is not questionable. However, our analysis underlines the fact that a direct comparison of the slope of the turbulence spectra with theoretical ones is not appropriate. In particular, we have shown that the flux of energy is not constant over the cascade, because the dissipation is relevant at each scale. Such mechanism is not yet taken into account in the WTT and further theoretical developments in that direction would be useful for our understanding of realistic wave turbulence dynamics.

\acknowledgments
The authors would like to thank Arezki Boudaoud for fruitful discussions about the energy budget of the cascade.\\
{\it\small $^\dag$ S.R. is on leave from Institut Non Lin\'eaire de Nice, UMR 6618 CNRS-UNSA - 1361 Route des Lucioles, 06560 Valbonne, France}

\end{document}